\documentclass[11pt]{article}
\usepackage{moriond,epsfig}

\def\be{\begin{equation}}
\def\ee{\end{equation}}
\def\bea{\begin{eqnarray}}
\def\eea{\end{eqnarray}}

\begin{document}
\vspace*{4cm}
\title{A GIANT AIR SHOWER ARRAY IN NORD RHEIN WESTFALEN}

\author{ HINRICH MEYER }

\address{Fachbereich Physik, Bergische Universit\"at Wuppertal, P.O. Box, \\
D-42097 Wuppertal, Germany}

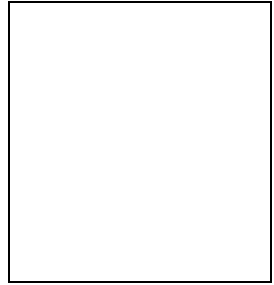
\maketitle\abstracts
{A proposal is presented to construct a very large air shower array in the
densely populated central area of Nord-Rhein-Westfalen (NRW). The aim would be
to reach a size of approximately 4000 km$^2$ to study the highest
energy end of the all particle cosmic ray spectrum, as it is 
accessible in the northern hemisphere.}

\section{Basic Considerations}       
The energy spectrum of cosmic rays appears to extend unattenuated 
beyond $5 \cdot 10^{19}\
eV$ although photoproduction of pions by collisions
of the primary protons with the universal $2.7^\circ\ K$ microwave
background should have cut off the cosmic ray energy spectrum\cite{ref1}. The
assumption that the highest energy particles of cosmic rays are protons
is a most natural one and is also in accord with the observed features
of the giant air showers. If the astrophysical sources of the cosmic
rays are on average far enough away, that is 100 Mpc or more and with
protons as the primary particles, the cut off
of the spectrum is unavoidable. Then we have the classical controversy
about distances of sources, near of far, that  played a fundamental role in
many astrophysical phenomena, like recently for the gamma ray bursts (GRB).
Only observations decide.

Two questions need to find a definite answer experimentally

\begin{itemize}
\item
do the highest energy particles come from preferred locations on the sky
(sources)
\item
what are the highest energies of cosmic rays (end of the spectrum)
\end{itemize}
and closely related to those two questions, what is the nature of the
particles: are they protons, nuclei, photons, neutrinos or rather exotic
objects.

The particle flux near the end of the spectrum is low, about one event per
100 km$^2$ and year. To collect sufficient statistics needed for a successful
experiment, a giant detector is required with more than 1000 km$^2$ area
to improve the acceptance by at least an order of magnitude over 
AGASA\cite{ref2}.

I have considered a highly populated and technically advanced area as a
viable solution to the requirements for a successful air shower array.
This follows the basic idea of AGASA but here for an area that is famous
for its high level of industrialisation. A look at the Rhein-Ruhr area 
between Duisburg, Dortmund, K\"oln and Bonn reveals that one will find a
power outlet almost every km. Furthermore, with D\"usseldorf, Essen,
Bochum, Wuppertal and a little farther out, M\"unster, Siegen and Aachen
(in addition to the 4 cities above) one
has an unparalleled density of universities. The triangle with Bonn, 
Duisburg and Dortmund at the corners covers an area of 4300 km$^2$ and
a triangle with Aachen, Siegen and M\"unster, an area twice as large can be
achieved for the deployment of a giant air shower array.

\section{Structure of the Array}

Considering a threshold of $10^{19}$ GeV one would choose, based on previous
experiments, about 1.5 km distance between stations. To efficiently cut
accidentals to a negligible level at a
given station, 3 detectors with spacing of 5 - 10 m are foreseen, 
operated
at a 2 out of 3 coincidence. The exact conditions can easily be chosen
such that a basic rate per station of a few Hz is obtained.

In addition to the amplitude vs. time from the detectors, an absolute time
derived from the global positioning system (GPS) is recorded for each
event. From a local buffer the information is transmitted via internet
to a local base preferably at the nearest university and the large
showers  are reconstructed offline. In this context it may be of considerable
interest for the project that power companies in NRW develop internet
connections via the power grid, the use of which then may considerably 
simplify the communication structure in the array.

The basic detector should be a 2 m$^2$ scintillator with wavelength
shifter readout using photodiodes. The detector should be covered by a
strong and tight box that can easily be located in any reasonable environment.

Of particular concern is the acceptance of particles at larger zenith
angles because of the possibility to catch highest energy neutrinos.
A recent reanalysis of Haverah Park data has shown the large potential
of events in the angular range $> 60^\circ$, in particular for primary particle
identification, in fact a rather strong upper limit on photons  as the 
highest energy particles could be derived\cite{ref3}.

The geometry of the scintillators should therefore allow for significant
acceptance at large zenith angles, which is certainly possible. The energy
threshold of a triple of the detectors at a given location will be around
100 TeV and from the three counters a crude angular pointing for each 
shower of about (5 -- 8)$^\circ$ can be derived. A large
array then allows to search for long range coincidences at scales of
10th of km which has been attempted previously with little success, but
also with insufficient instrumentation. A large array then with several 
thousand stations would certainly be of considerable value to
investigate this problem.

As a further strategy point I would prefer to go for the largest 
possible area, even at the expense of the density of stations at the
early phase of
the experiment. This will shift up the energy threshold, however
to detect events far above the GZK cutoff would be of 
great importance and is one of the main goals of the experiment.

In line with the discussion above one should start with part of the array
in a rather dense environment that means at the center of one of the larger
cities. As an example I have chosen (of course) Wuppertal,
which is located approximately in the center of the 4300 km$^2$ triangle.
Within the city limit an area of about 240 km$^2$ is enclosed. Dividing
the area up into 2 km $\times$ 2 km squares 60 stations need to be 
deployed. This results in an array that is 2.4 times larger than
AGASA which would provide a very reasonable start.
The cities of Duisburg, Dortmund and Bonn would mark the corners of the
triangle and should be instrumented in a similar way. 

\section{Public Outreach}

A further important consideration concerns the type of locations for the
deployment of the stations. It appears that e.g. for the case of
Wuppertal a considerable fraction of the 2 km $\times$ 2 km squares contain
one or more schools. It then seems prudent to consider them as the preferred
location for the stations. The schools are for the large majority on
public ground. They have power and can be considered a well protected place.
In addition, both teachers and pupils should develop interest to
participate in the experiment as such. No school would be preferred, each of
the stations is of exactly the same importance in the array. This way 
the schools will be an integral part of an
experiment in basic science, connected by a very large
network in the central part of NRW.

Actually, the network may as well form the basis for other research
activities that require a 100\% ontime in a very well controlled manner.
I think one should not underestimate the possible impact of such an 
enterprise for  popularising basic science in schools and for building
and maintaining connections beween schools and universities.

\section{Conclusion}

The construction of a giant air shower array in the central part of
NRW is proposed. The area of 4300 km$^2$ foreseen makes it comparable
in size to the AUGER projects. Unique of this proposal is, however,
the connection to public schools that provides an elegant opportunity of
popularising basic science.

\section*{Acknowledgments}

I would like to thank the organisers of this meeting for providing a very
stimulating atmosphere and for a generous support.

\end{document}